# First Identification of New X-Ray Spectra of $Mo^{39+}$, $Mo^{40+}$, $W^{43+}$, $W^{44+}$ and $W^{45+}$ on EAST

Fudi Wang[1,*], Dian Lu[1,2], Mingfeng Gu[3], Yifei Jin[1,2], Jia Fu[1], Yuejiang Shi[4,*], Yang Yang[5,6], J. E. Rice[7], Manfred Bitter[8], Qing Zang[1], Hailin Zhao[1], Liang He[1,2], Miaohui Li[1], Handong Xu[1], Haijing Liu[9], Zichao Lin[1,2], Yifei Chen[10], Yongcai Shen[10], Kenneth Hill[8], Cheonho Bae[1], Shengyu Fu[9], Hongming Zhang[1], Sanggon Lee[11], Xiaoqing Yang[1], Guozhang Jia[1], Yingying Li[4], Bo Lyu[1], Juan Huang[1], Xianzu Gong[1] and Baonian Wan[1]

[1] *Institute of Plasma Physics, HFIPS, Chinese Academy of Science, Hefei 230031, China*
[2] *Science Island Branch, Graduate School of University of Science and Technology of China, Hefei 230031, China*
[3] *SSL, U.C. Berkeley, CA-94720, United States of America*
[4] *Enn Science and Technology Development Co., Ltd, Langfang, China*
[5] *Shanghai EBIT Laboratory, Institute of Modern Physics, Fudan University, Shanghai 200433, China*
[6] *Key Laboratory of Nuclear Physics and Ion-beam Application (MOE), Shanghai 200433, China*
[7] *PSFC, MIT, Cambridge, MA-02139, United States of America*
[8] *Princeton Plasma Physics Laboratory, MS37-B332, Princeton, NJ 08543-0451, USA*
[9] *School of Electrical Engineering, University of south China, Hengyang 421001, China*
[10] *School of Physics and Materials Engineering, Hefei Normal University, Hefei 230601, China*
[11] *National Fusion Research Institute, 52 Eoeun-Dong, Yusung-Gu, Daejeon 305-333, Korea*

New high-resolution x-ray spectra of $Mo^{39+}$, $Mo^{40+}$, $W^{43+}$, $W^{44+}$ and $W^{45+}$ have been carefully confirmed for the first time by use of the x-ray imaging crystal spectrometer (XCS) in Experimental Advanced Superconducting Tokamak (EAST) under various combined auxiliary heating plasmas conditions. Wavelength of these new x-ray spectra is ranged from 3.895 Å to 3.986 Å. When core electron temperature ($T_{e0}$) reaches 6.0 keV, $Mo^{39+}$ and $Mo^{40+}$ lines of 3.9727, 3.9294 and 3.9480 Å can be effectively detected on XCS for EAST; meanwhile, line-integrated brightness of these spectral lines of $Mo^{39+}$ and $Mo^{40+}$ is very considerable when electron temperature reaches 12.9 keV. Multi-components spectral lines for $W^{43+}$, $W^{44+}$ and $W^{45+}$ have also been identified when $T_{e0}$ reaches 6 keV. Parts of spectral lines, such as Zn-1, Cu-2, Cu-4a, Cu-4d and Cu-5 lines of tungsten, are first observed experimentally. When electron temperature reaches 12.9 keV, line-integrated intensity for part of these spectral lines of $W^{43+}$, $W^{44+}$ and $W^{45+}$ are considerable. These experimental results and theoretical predictions from FAC and FLYCHK codes are in good general agreement. These new spectral lines, obtained on XCS for EAST, are vital for deeply uncovering the mechanisms of ion and electron thermal, high-Z impurity and momentum (anomalous) transport to achieve the advanced steady-state operation scenarios for the present tokamaks, ITER and CFETR.

Author to whom correspondence should be addressed: fdwang@ipp.ac.cn and yjshi@ipp.ac.cn

Understanding the ion and electron thermal, particle, high-Z impurity and momentum (anomalous) transport, driven by the ITG, TEM and ETG drift wave turbulence, is recognized as one of the most important issues for the present tokamaks and future Fusion Experimental Reactors, in order to obtain the advanced steady-state operation scenarios with the high level of energy confinement, wider internal transport barrier, and strongly reversed magnetic shear [1-3]. Generally, the drift wave turbulence is driven by the different free energy for micro-instabilities, mainly the gradient of plasma temperature, density and the equilibrium magnetic field. So the temperature and density profiles are crucial parameters for ITER and CFETR.

High resolution x-ray spectroscopy has made significant contributions to the diagnostics of plasmas by providing data on the ion and electron temperature, impurity ion transport, plasma rotation, impurity ion-charge state distributions [4-30]. In general, these data are derived from the spectra of medium- and high-Z elements like Si, Ar, Ti, Cr, Fe, Ni, Mo, Xe and W which exist in tokamak and stellarator plasmas as indigenous impurities or which are injected as trace elements by a laser blow off system or gas puffing.

In EAST plasmas, tungsten and molybdenum ions are indigenous impurities because of the successive use of the upper and lower tungsten divertors [31] and the use of molybdenum tiles on the wall and divertor surfaces [32-34]. Additionally, the flexible atomic code (FAC) [35-39] and FLYCHK code [40-42] also predicts the existence of x-rays and the fractional abundance of tungsten and molybdenum ions [43]. However, some of the x-ray lines of the tungsten and molybdenum ions, which are predicted by the FAC and FLYCHK codes, have, so far, not yet been observed from tokamaks and stellarators plasmas.

The present paper reports new experimental results from recent observations of the x-ray spectra of $Mo^{39+}$, $Mo^{40+}$, $W^{43+}$, $W^{44+}$ and $W^{45+}$ from EAST tokamak plasmas with very high electron temperatures in the range from 6 to 16 keV. These experimental results are in good agreement with the predictions from the FAC and FLYCHK codes. These new spectra are crucial for the investigations of the ion and electron thermal, the tungsten and molybdenum ions, toroidal and poloidal momentum transport on EAST and future Fusion Experimental Reactor, such as ITER and CFETR, etc.

The experimental parameters on EAST are as follows: EAST is a full superconducting tokamak with a non-circular plasma cross section. The typical major and minor radii are $R = 1.85$ m and $a = 0.45$ m, respectively. The operating range of the toroidal magnetic field ($B_T$) is from 2.0 to 3.5 Telsa at $R = 1.70$ m. The plasma current (Ip) is in the range from 0.25 to 1 MA. The typical line-average electron density ($n_e$) is in the range from 1.6 to $6 \times 10^{19}$/m$^3$. EAST has the flexible (diverted and limited) magnetic configurations and advanced lithium wall conditioning techniques to reduce the particle recycling, and is optimized for long-pulse high-performance plasma operations with upgraded auxiliary heating and current drive systems and upper and lower tungsten divertors for the handling high-power exhaust.

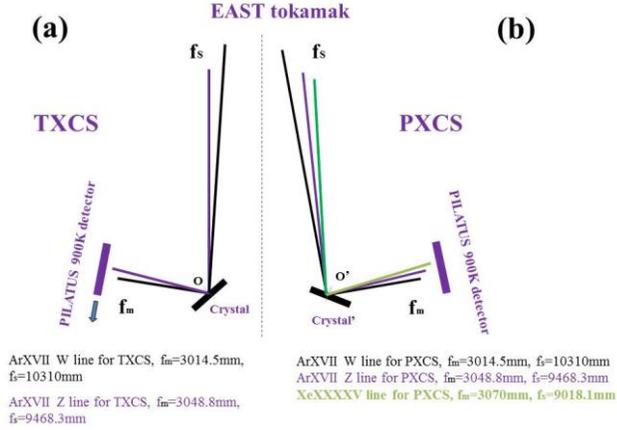

FIG. 1 (a) Top view of meridional focal points ($f_m$) and sagittal focal points ($f_s$) for W and Z line of ArXVII on TXCS; (b) Top view of $f_m$ and $f_s$ for XeXLV line and ArXVII's W and Z line on PXCS.

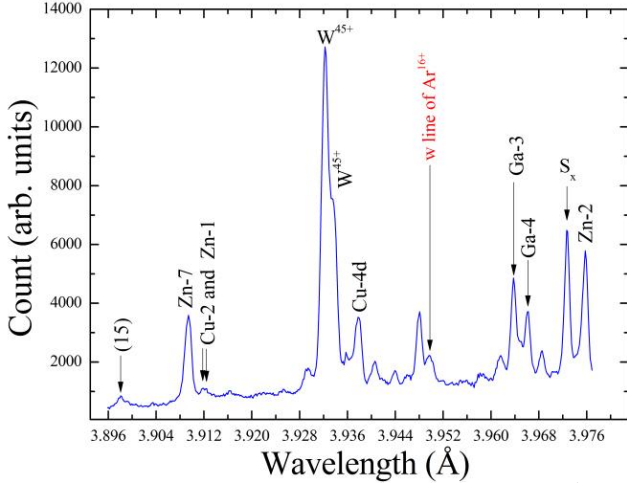

FIG. 2 The spectrum in the wavelength range from 3.895 to 3.978 Å from spatial-integrated of 5cm on mid-plane plasmas in shot No. 98958. Integration time is from 29.8 to 30.2 s. Denoted individual spectral lines features are described in Table 1.

During the period from 2010 to 2012, the graphite tiles on the wall and divertor surfaces – except for the divertor target plates – were replaced molybdenum tiles in order to ensure density control [44]; and during the period from 2014 to 2021, the upper and lower divertors of EAST were continuously upgraded to tungsten divertors [31]. With respect to auxiliary heating, EAST is equipped with an electron cyclotron resonance heating (ECRH) system [45-47] with a nominal power of 3 MW at 140 GHz, two lower hybrid current drive (LHCD) systems [48-51] with total rated power of 10 MW at 2.45 and 4.6 GHz, an ion cyclotron range of frequency (ICRF) heating system [52] with a rated power of 12 MW in the wide range of frequency 25 – 70 MHz, and two neutral beam injection (NBI) systems [53, 54] with a maximum power of 8 MW.

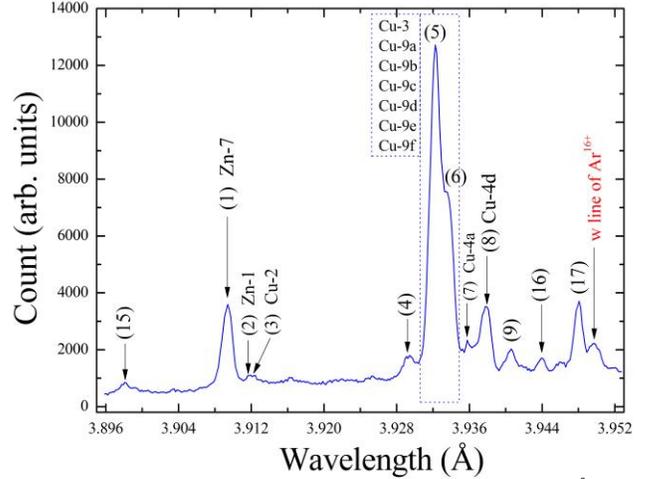

FIG. 3 The spectrum in the wavelength range from 3.895 to 3.953 Å from spatial-integrated of 5cm on mid-plane plasmas in shot No. 98958. Integration time is from 29.8 to 30.2 s. Denoted individual spectral lines features are described in Table 1.

The experimental results, presented in this letter, were obtained during the latest EAST experimental campaigns (2018.03 – 2022.01) from tokamak plasmas with varying combinations of auxiliary heating, which comprised the combined heating of LHCD and ECRH, the combined heating LHCD, ECRH and ICRF, the combined heating of LHCD, ECRH and NBI, the combined heating of LHCD, ECRH, ICRF and NBI. The electron temperature, in the plasma region with $\rho \sim 0 - 1$, was measured with the electron cyclotron emission (ECE) [55, 56] and Thomson scattering (TS) [57] systems. The x-ray spectra and line-integrated brightnesses of $Mo^{39+}$, $Mo^{40+}$, $W^{43+}$, $W^{44+}$ and $W^{45+}$ were recorded by the x-ray imaging crystal spectrometers (XCS) [27, 28, 58, 59]. In order to effectively and stably obtain these x-ray spectra of $Mo^{39+}$, $Mo^{40+}$, $W^{43+}$, $W^{44+}$ and $W^{45+}$, by utilizing the novel PILATUS 900K detector design, which was installed on a rail driven by a linear slider [59], the center position of PILATUS 900K detector on XCS can be flexibly moved to the wavelength position of interest, such as, the position of short wavelength, which was close to w-line of ArXVII and corresponded to the Bragg angle of 53.5004$^0$ - *see FIG. 1*. The wavelength calibration is based on comparisons with the theoretical wavelengths for the ArXVII ions [60] which fall in the observable spectral range of XCS. The uncertainty for the wavelength measurements is ± 0.2 mÅ. Relative molybdenum and tungsten ion population, in different electron temperature, was calculated by the FLYCHK code [40-42], which solves the rate equations for level population distributions by considering collisional and radiative atomic processes. Theoretical wavelengths of molybdenum and tungsten x-ray spectra were obtained by the flexible atomic code (FAC) [35-39], which applies a fully relativistic approach based on the Dirac equation and is capable of treating direct collisional excitation, radiative transition, radiative recombination, ionization by electron impact nonresonant photoionization, dielectronic recombination and autoionization.

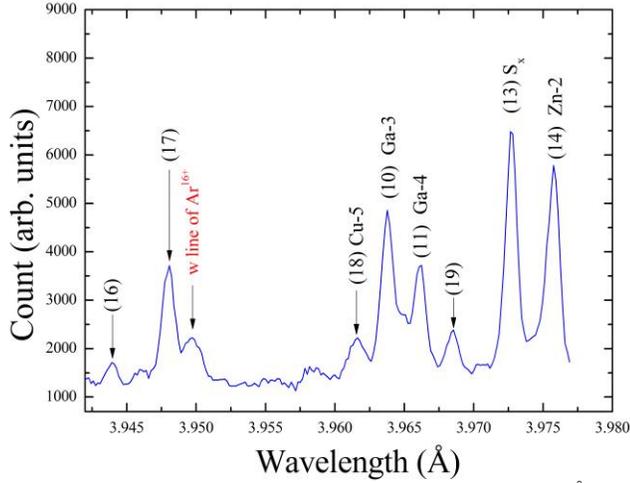

FIG. 4 The spectrum in the wavelength range from 3.942 to 3.978 Å from spatial-integrated of 5cm on mid-plane plasmas in shot No. 98958. Integration time is from 29.8 to 30.2 s. Denoted individual spectral lines features are described in Table 1.

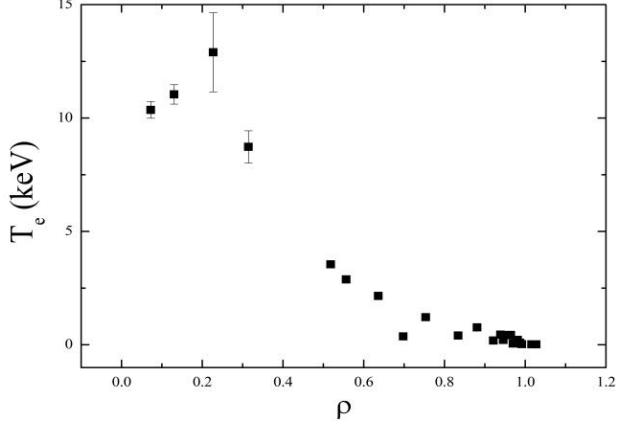

FIG. 5 Electron temperature from TS in shot No. 98958.

FIG. 2 – 4 show the x-ray spectra of $Mo^{39+}$, $Mo^{40+}$, $W^{43+}$, $W^{44+}$, $W^{45+}$ and $Ar^{16+}$ from a tokamak plasma with LHCD and ECRH (*EAST shot No.98958*), where the maximum electron temperature was 12.9 keV. The labeled spectral features are explained in Table 1. The experimental parameters for *EAST shot No.98958* were as follows: The plasma current was $I_p$ = 0.5 MA in the counterclockwise direction from top view of EAST. The plasma current Ip was fully driven by LHCD and ECRH since the loop voltage was $V_{loop}$ = 0 V. The toroidal magnetic field was $B_T$ = 2.75 T at $R$ = 1.70 m in the clockwise direction as seen from the top view of EAST. During the injection phase of LHCD and ECRH, the central line average electron density was $n_e$ = 1.75 $\times 10^{19}$/m$^3$ and the plasma configuration was a lower single null configuration.

The electron temperature profile obtained from TS is shown in FIG. 5; the discharge obviously had an electron internal transport barrier (eITB) in the region of $\rho$ ~ 0.2 – 0.5. The maximum electron temperature was 12.9 keV. The plasma stored energy ($W_{MHD}$), $\beta_N$ and $\beta_P$ were 112 – 122 kJ, 0.667 – 0.8 and 0.684 – 0.844, respectively. $q_0$, $q_{95}$ and $l_i$ were 1.04 – 1.10, 5.40 – 5.46 and 1.42 – 1.50, respectively. The injection power of LHCD2 at 4.6 GHz with an $n_\parallel$ of 2.26 on window E was 2.35 MW; and the injection period of LHCD2 was from 1.85 to 103.5 s. The injection power for ECRH1, ECRH2 and ECRH3 at 140 GHz on window M was 0.5 MW, 0.45 MW and 0.4 MW, respectively. The injection period of ECRH1 and ECRH3 was from 2 s to 104 s and the injection period of ECRH2 was from 2.4 s to 104.4 s. The toroidal and poloidal angles for both, the ECRH1 and the ECRH3, were $200^0$ and $103^0$, respectively; whereas the toroidal and poloidal angles for ECRH2 were $180^0$ and $75^0$, respectively.

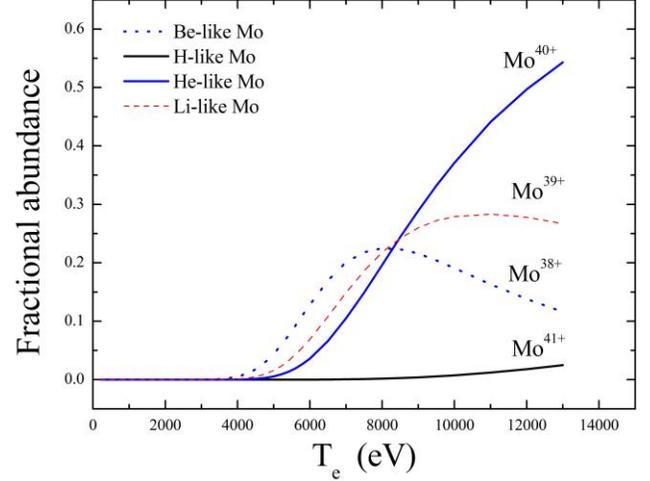

FIG. 6 Relative molybdenum ion population in different electron temperature.

The $Mo^{39+}$ ($S_x$, $1s^22p$ $^1P^0_{1/2}$ – $1s^23s$ $^2S_{1/2}$, 3.9727 Å) spectral line was identified with a high level of confidence: Firstly, the experimental wavelength of $S_x$ on XCS is 3.9727 Å and the theoretical wavelength of $1s^22p$ $^1P^0_{1/2}$ – $1s^23s$ $^2S_{1/2}$ of $Mo^{39+}$ is 3.97263 Å [61, 62]. The values for these two wavelengths are in good agreement. Secondly, when the electron temperature, shown in FIG. 5, is 12.9 keV, the line-integrated brightness of $S_x$ ($S_x$, $1s^22p$ $^1P^0_{1/2}$ – $1s^23s$ $^2S_{1/2}$, 3.9727 Å), shown in FIG. 2 and FIG. 4, is very strong (maximum line-integrated brightness of 6480 a.u. on PILATUS 900K detector). But when electron temperature, shown in FIG. 10, is 6.0 keV, the line-integrated brightness of $S_x$, shown in FIG. 7 and FIG. 9, is extremely weak and the line almost disappears. The experimental observation results are also in good agreement with the predictions of FLYCHK [40-42] shown in FIG. 6. As shown in FIG. 6, the relative population of $Mo^{39+}$ is very low and about 6%, when electron temperature is 6.0 keV. But when the electron temperature is 12.9 keV, the relative population of $Mo^{39+}$ reaches a value of 27%; more notably, the relative population of $Mo^{39+}$ reaches the maximum value of 28.3% in the plasma region of $\rho$ = 0.15 where the electron temperature is 11 keV – *see FIG. 5 and FIG. 6*. Thirdly, the line-integrated ion temperatures inferred from $Mo^{39+}$ $S_x$ line and the w-line of $Ar^{16+}$ are 0.796 ±0.082 keV and 0.69 ±0.31 keV, respectively. The ion temperature of two kinds of ions is very close. It is noted that the ion temperature of $Mo^{39+}$ is slightly higher than the ion temperature of $Ar^{16+}$ due to the fact that the $Mo^{39+}$ is much closer to the plasmas core region and more abundant than $Ar^{16+}$ ions when the electron temperature is 12.9 keV. Fourthly, the $Mo^{39+}$ ion obviously originates from the molybdenum tiles on the wall and divertor surfaces in EAST [44].

Table 1 Tungsten, molybdenum and ArXVII spectrum in the wavelength range from 3.895 Å to 3.986 Å according to Ref. [43, 60-65] and NIST Atomic Spectra Database.

| $\lambda_{ex}$ (Å) | $\lambda_{th}$ (Å) | Ion | Lab. | Transition designation |
|---|---|---|---|---|
| 3.8980 | ? | ? | (15) | ? |
| 3.9095 | 3.9098 | $W^{44+}$ | Zn-7 | $3d^{10}4s^2\ ^1S_0 - 3d^9(^2D_{3/2})4s^26f(3/2,5/2)^0\ 1$ |
| 3.9116 | 3.9115 | $W^{44+}$ | Zn-1 | $3d^{10}4s^2\ ^1S_0 - 3d^94s^26f\ ^1P_1$ |
| 3.9121 | 3.9118 | $W^{45+}$ | Cu-2 | $3d^{10}4s^2S_{1/2} - 3d^94s6g^2F_{5/2}$ |
| 3.9294 | 3.9312 | $Mo^{40+}$ | (4) | $1s2p\ ^3P^0_2 - 1s3s\ ^3S_1$ |
| Y | 3.9308 | $W^{45+}$ | Cu-3 | $3d^{10}4s^2S_{1/2} - 3d^94s5d^2P_{3/2}$ |
| 3.9329 | 3.9328 | $W^{45+}$ | Cu-9a | $3d^{10}4s^2S_{1/2} - 3d^94s\ (5/2,1/2)_2 6f_{7/2}\ (2,7/2)^0\ 3/2$ |
| 3.9329 | 3.9328 | $W^{45+}$ | Cu-9b | $3d^{10}4s^2S_{1/2} - 3d^94s\ (5/2,1/2)_3 6f_{7/2}\ (3,7/2)^0\ 1/2$ |
| 3.933 | 3.9330 | $W^{45+}$ | Cu-9c | $3d^{10}4p^2P^0_{1/2} - 3d^94p\ (5/2,1/2)^0_2 6f_{7/2}\ (2,7/2)\ 3/2$ |
| 3.933 | 3.9330 | $W^{45+}$ | Cu-9d | $3d^{10}4p^2P^0_{1/2} - 3p^53d^{10}4p\ (3/2,1/2)_2 5d_{5/2}(2,5/2)\ 3/2$ |
| 3.933 | 3.9330 | $W^{45+}$ | Cu-9e | $3d^{10}4s^2S_{1/2} - 3p^53d^{10}4s\ (3/2,1/2)^0_2\ 5d_{5/2}\ (2,5/2)^0\ 1/2$ |
| 3.933 | 3.9330 | $W^{45+}$ | Cu-9f | $3d^{10}4s^2S_{1/2} - 3p^53d^{10}4s\ (3/2,1/2)^0_2\ 5d_{5/2}\ (2,5/2)^0\ 3/2$ |
| 3.9357 | 3.9348 | $W^{45+}$ | Cu-4a | $3d^{10}4s^2S_{1/2} - 3d^94s5d^2D_{3/2}$ |
| Y | 3.9364 | $W^{45+}$ | Cu-4b | $3d^{10}4s^2S_{1/2} - 3d^94s5d^2P_{1/2}$ |
| Y | 3.9364 | $W^{45+}$ | Cu-4c | $3d^{10}4s^2S_{1/2} - 3d^94s6f^2P_{1/2}$ |
| 3.9378 | 3.9368 | $W^{45+}$ | Cu-4d | $3d^{10}4s^2S_{1/2} - 3d^94s6f^2P_{3/2}$ |
| 3.9406 | ? | ? | (9) | ? |
| 3.9439 | ? | ? | (16) | ? |
| 3.9480 | 3.9461 | $Mo^{40+}$ | (17) | $1s2p\ ^1P^0_2 - 1s3s\ ^1S_0$ |
| 3.9494 | 3.9493 | $Ar^{16+}$ | w | $1s^2\ ^1S_0 - 1s2p\ ^1P_1$ |
| 3.9616 | 3.9621 | $W^{45+}$ | Cu-5 | $3p^{16}4p^2P_{1/2} - 3p^54s5f^2D_{3/2}$ |
| 3.9637 | 3.9636 | $W^{43+}$ | Ga-3 | $3d^{10}4p^2P^0_{1/2} - 3d^94p6f^2\ D_{3/2}$ |
| 3.9660 | 3.9661 | $W^{43+}$ | Ga-4 | $3d^{10}4p^2P^0_{1/2} - 3d^94p6f^2\ P_{1/2}$ |
| 3.9685 | ? | ? | (19) | ? |
| 3.9698 | 3.9696 | $Ar^{16+}$ | y | $1s^2\ ^1S_0 - 1s2p\ ^3P_1$ |
| 3.9727 | 3.9726 | $Mo^{39+}$ | $S_x$ | $1s^22p\ ^1P^0_{1/2} - 1s^23s\ ^2S_{1/2}$ |
| 3.9757 | 3.9734 | $W^{44+}$ | Zn-2 | $3p^64s^2\ ^1S_0 - 3p^54s^25d\ ^1P_1$ |

Also, the $Mo^{40+}$ spectral line ((17), $1s2p\ ^1P^0_2 - 1s3s\ ^1S_0$, 3.9480 Å) and $Mo^{40+}$ spectral line ((4), $1s2p\ ^3P^0_2 - 1s3s\ ^3S_1$, 3.9294 Å) were carefully identified, following the above procedure of studying the behavior of these features under different conditions (see FIG. 2, FIG. 3, FIG. 4, FIG. 5, FIG. 6, FIG. 7, FIG. 8, FIG. 10 and Table 1). We point out that the experimental results, shown in FIG. 2, FIG. 3, FIG. 4, FIG. 7, FIG. 8, FIG. 9 and theoretically defined in Table 1, represent the first observations and clear identification of the following three spectral lines:

$Mo^{39+}$ ($S_x$, $1s^22p\ ^1P^0_{1/2} - 1s^23s\ ^2S_{1/2}$, 3.9727 Å),
$Mo^{40+}$ ((17), $1s2p\ ^1P^0_2 - 1s3s\ ^1S_0$, 3.9480 Å)
$Mo^{40+}$ ((4), $1s2p\ ^3P^0_2 - 1s3s\ ^3S_1$, 3.9294 Å)

And we also point out that $Mo^{40+}$ is the highest molybdenum charge state observed in tokamak plasmas, surpassing $Mo^{37+}$ previously observed [43].

FIG. 7 – 9 show the typical x-ray spectra of $Mo^{39+}$, $Mo^{40+}$, $W^{43+}$, $W^{44+}$, $W^{45+}$ and $Ar^{16+}$ from a plasma with LHCD and ECRH and a maximum electron temperature of 6.0 keV produced in shot No.107006 on EAST. The labeled spectral lines features are explained in Table 1. The experimental parameters for shot No.107006 were as follows: The plasma current was Ip = 0.4 MA in the counterclockwise direction seen from top view of EAST. The loop voltage was $V_{loop}$ = 0 V, so that Ip was fully driven by the LHCD and ECRH. The toroidal magnetic field was $B_T$ = 2.75 T at R = 1.70 m and the direction of $B_T$ was in the counterclockwise direction seen from top view of EAST. During the injection phase of LHCD and ECRH, the central line average electron density was $n_e$ = 1.80 × $10^{19}$/m$^3$ and the plasma configuration was double null configuration with elongation of about 1.70.

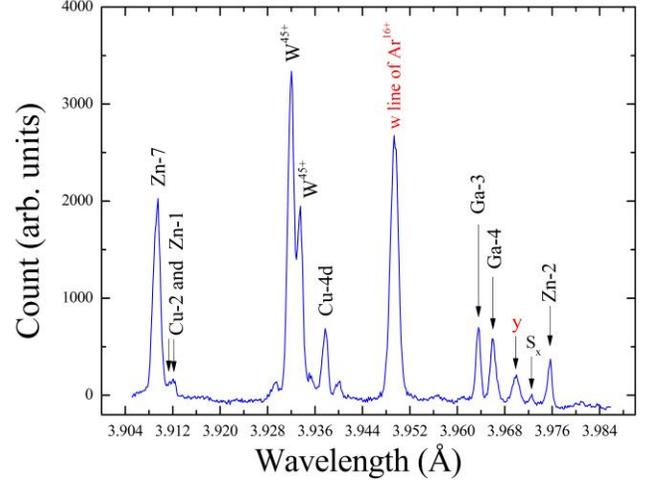

FIG. 7 Emission spectrum in the 3.904 − 3.986 Å range from spatial-integrated of 5cm on mid-plane plasmas in shot No. 107006. Integration time is from 4.8 to 5.2 s. The Denoted individual emission features are described in Table 1.

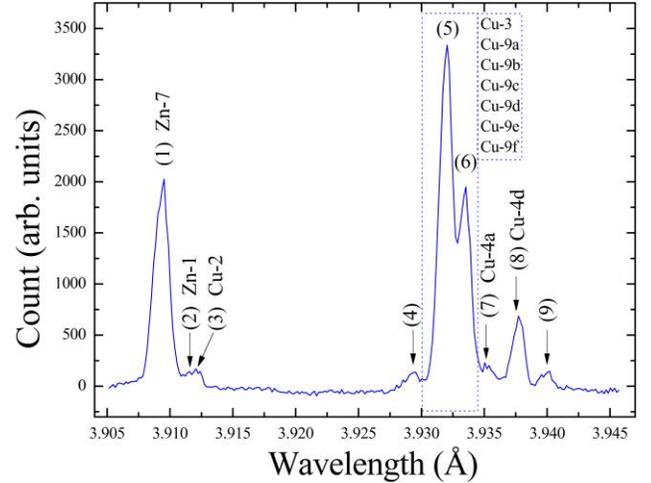

FIG. 8 The spectrum in the wavelength range from 3.904 to 3.9457 Å from spatial-integrated of 5cm on mid-plane plasmas in shot No. 107006. Integration time is from 4.8 to 5.2 s. Denoted individual spectral lines features are described in Table 1.

The electron temperature at 4.0, 5.0 and 6.0 s obtained from ECE is shown in FIG. 10 and the maximum electron temperature was 6.0 keV. $W_{MHD}$, $\beta_N$ and $\beta_P$ were 103 – 112 kJ, 1.17 – 1.27 and 0.809 – 0.87, respectively. $q_0$, $q_{95}$ and $l_i$ were 1.489 – 1.675, 7.6 – 7.823 and 1.365 – 1.475, respectively. The injection power of LHCD1 at 2.45GHz with an $n_\parallel$ of 2.84 on window N and the injection power of LHCD2 at 4.6 GHz with an $n_\parallel$ of 2.04 on window E were 0.4 MW and 1.4 MW, respectively. The injection periods for LHCD1 and LHCD2 were from 4.97 to 10.7 s and from 1.7 to 9.3 s, respectively. The injection power of ECRH1 and ECRH3 at 140 GHz on window M were 0.6 MW and 0.5 MW, respectively. The injection period of ECRH1 and ECRH3 was from 1.46 to 9.88 s. The toroidal and poloidal angles of ECRH1 were $200^0$ and $77^0$, respectively; whereas the toroidal and poloidal angles of ECRH3 were $200^0$ and $103^0$, respectively.

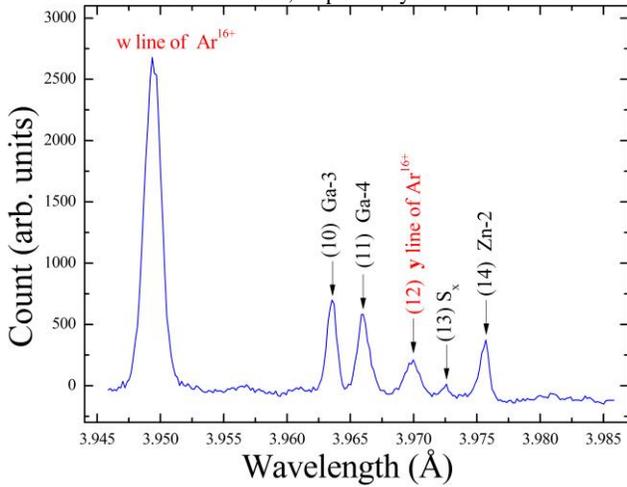

FIG. 9 The spectrum in the 3.9459—3.986 Å range from spatial-integrated of 5cm on mid-plane plasmas in shot No. 107006. Integration time is from 4.8 to 5.2 s. Denoted individual spectrum features are described in Table 1.

The $W^{44+}$ (Zn-7, $3d^{10}4s^2\ ^1S_0 - 3d^9(^2D_{3/2})4s^26f(3/2,5/2)^0$ 1, 3.9095 Å) spectral line was carefully identified: Firstly, the experimental and theoretical wavelengths of Zn-7 are 3.9095 Å and 3.9098 Å, respectively. The values of these two wavelengths are in good agreement. Secondly, $T_{e0}$, shown in FIG. 10, is 6.0 keV and the line-integrated brightness of Zn-7, shown in FIG. 7 and FIG. 8, are considerable (maximum line-integrated brightness of 2027 a.u. on PILATUS 900K detector). But when electron temperature is about 1.2 keV, the line-integrated intensity of the Zn-7 line is very weak. The experimental results are in good agreement with prediction of FLYCHK [40-42] shown in FIG. 11. As shown in FIG. 11, when electron temperature is 1.2 keV, the relative population of $W^{44+}$ is very low and about 4.0%. But when the electron temperature in the plasma region of $\rho = 0.50$ reaches 2.0 keV ($T_{e0} = 6.0$ keV), the relative population of $W^{44+}$ reaches the maximum value of 24.4% – see FIG. 10 and FIG. 11. Thirdly, the $W^{44+}$ ion obviously originates from the upper and lower tungsten divertors in EAST [31]. Fourthly, no spectral lines of tungsten were detected by the XCS from discharges without the use of tungsten divertors, when $T_{e0}$ reached 2.2 keV during the 2012 - 2014 EAST experimental campaigns [66]. We point out that our data present the first clear identification of the Zn-7 line of tungsten from tokamak plasmas.

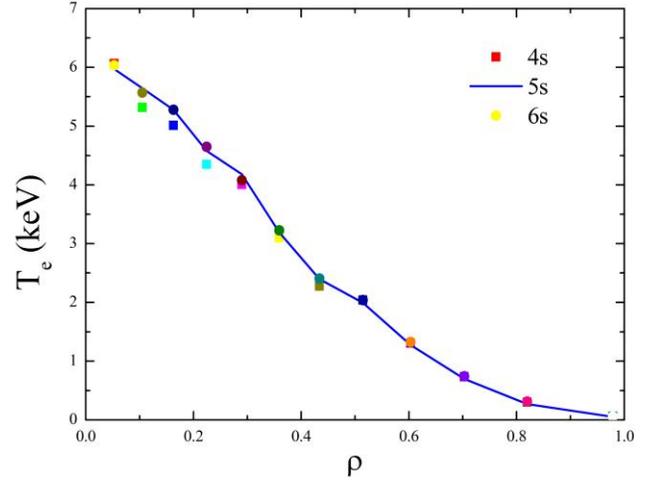

FIG. 10 Electron temperature from ECE at 4.0, 5.0 and 6.0 s in shot No. 107006.

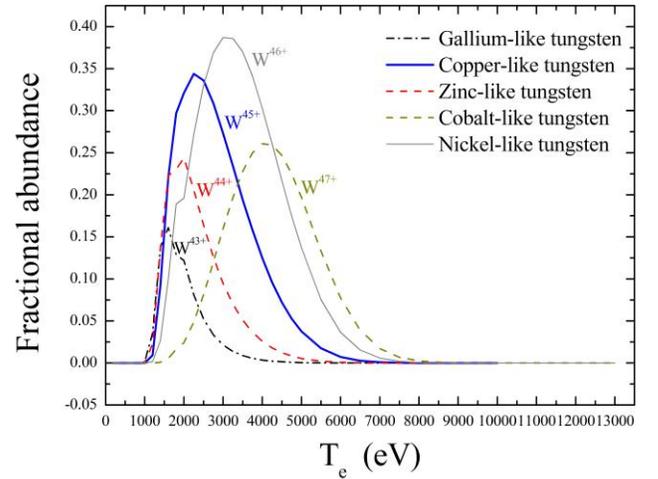

FIG. 11 Relative tungsten ion population in different electron temperature.

Also, the Cu-3, Cu-9a, Cu-9b, Cu-9c, Cu-9d, Cu-9e and Cu-9f lines of tungsten are identified, following the above procedure of investigating the behavior of these features under different conditions (see FIG. 2, FIG. 3, FIG. 7, FIG. 8, FIG. 10 and Table 1), but these lines cannot be distinguished one by one due to the insufficient dispersion of XCS. The Ga-3, Ga-4 and Zn-2 lines of tungsten are also carefully identified, in accordance with the above procedure of studying the behavior of these features under different conditions (see FIG. 2, FIG. 4, FIG. 7, FIG. 9, FIG. 10 and Table 1). The three lines are also discussed in the Ref. [43]. In addition, the Zn-1 and Cu-2 lines of tungsten are also carefully identified, in the light of the above identification process of researching the behavior of these features under different conditions (See FIG. 2, FIG. 3, FIG. 7, FIG. 8, FIG. 10 and Table 1), whereas the two lines cannot be distinguished one by one because of the insufficient dispersion of XCS. We point out that the Zn-1 and Cu-2 lines of tungsten are first observed and identified experimentally, and theoretical wavelengths of these two lines are calculated in the Table 1 of the Ref. [43] by use of the FAC code[35-39].

Moreover, the Cu-4a, Cu-4d and Cu-5 lines of tungsten are also identified, according to above identification logic of studying the behavior of these features under different conditions (see FIG. 2, FIG. 3, FIG. 4, FIG. 7, FIG. 8, FIG. 10 and Table 1). We also point out that the three lines are first observed experimentally, and theoretical wavelengths of the three lines are also calculated in the Table 1 of Ref. [43] by use of the FAC code [35-39]. The $Ar^{16+}$ resonance (w, $1s^2$ $^1S_0$–$1s2p$ $^1P_1$, 3.9494Å) and the $Ar^{16+}$ intercombination (y, $1s^2$ $^1S_0$ −$1s2p$ $^3P_1$, 3.9696 Å) lines are easily identified in the plasmas with the low $T_{e0}$ of 1.2 keV on EAST. The theoretical wavelengths of the Cu-4b and Cu-4c lines of tungsten, which are from the FAC code, are both 3.9364 Å shown in the Table 1 of the Ref. [43]. The two lines are difficult to distinguish one by one owing to the insufficient dispersion of XCS. Unfortunately, according to the current understanding, it is difficult to identify the individual spectral lines features of the four lines which are the ((9), 3.9406 Å), ((16), 3.9439 Å), ((19), 3.9685 Å) and ((15), 3.8980 Å) spectral lines, shown in the FIG. 2, FIG. 3, FIG. 4, FIG. 7, FIG. 8 and Table 1.

In summary, new high-resolution x-ray spectra of $Mo^{39+}$, $Mo^{40+}$, $W^{43+}$, $W^{44+}$ and $W^{45+}$ have been carefully identified by use of XCS, utilizing the novel design for PILATUS 900K which was installed on a rail driven by a linear slider, in EAST under various combined auxiliary heating plasmas conditions. When electron temperature is 6.0 – 16 keV, $Mo^{39+}$ and $Mo^{40+}$ lines of 3.9727, 3.9294 and 3.9480 Å can be effectively detected. Multi-components spectral lines for $W^{43+}$, $W^{44+}$ and $W^{45+}$ have also been identified when electron temperature is 6 – 16 keV. Parts of spectral lines, such as Zn-1, Cu-2, Cu-4a, Cu-4d and Cu-5 lines of tungsten, are first observed experimentally. These experimental results and theoretical predictions from the FAC and FLYCHK codes are in good general agreement. These new spectral lines, recorded by the XCS on EAST, are very important for measurements of the ion and electron temperature, high-Z impurity, rotation velocity for present tokamaks and the future Fusion Experimental Reactor. Furthermore, these new spectral lines are also crucial for uncovering the mechanisms of ion and electron thermal, high-Z impurity and momentum (anomalous) transport, in order to obtain the advanced steady-state operation scenarios.

The authors are grateful to the ECRH group, LHCD group, diagnostic group, vacuum group, ICRF group, NBI group and operation team of EAST. The work is supported by National Magnetic Confinement Fusion Science Program of China (2017YFE0301300, 2019YFE03040000, 2019YFE03040003), National Natural Science Foundation of China (12175278), Anhui Provincial Natural Science Foundation (1908085J01), Comprehensive Research Facility for Fusion Technology Program of China (2018-000052-73-01-001228), Open Fund of Magnetic Confinement Fusion Laboratory of Anhui Province (2021AMF01002) and CAS President's International Fellowship Initiative (2022VMB0007), Anhui Provincial Natural Science Foundation (2008085QA39), The University Synergy Innovation Program of Anhui Province (GXXT-2021-029), Anhui Provence key research and development program (No. 202104a06020021) and ASIPP Science and Research Grant (No. DSJJ-2020-02).

[1] E.J. Doyle, W.A. Houlberg, Y. Kamada, et al., Nucl Fusion, 47 (2007) S18-S127.
[2] M. Wakatani, V.S. Mukhovatov, K.H. Burrell, et al., Nucl Fusion, 39 (1999) 2175-2249.
[3] C. Gormezano, A.C.C. Sips, T.C. Luce, et al., Nucl Fusion, 47 (2007) S285-S336.
[4] M. Bitter, S. Vongoeler, R. Horton, et al., Phys Rev Lett, 42 (1979) 304-307.
[5] M. Bitter, K.W. Hill, N.R. Sauthoff, et al., Phys Rev Lett, 43 (1979) 129-132.
[6] K.W. Hill, S.V. Goeler, M. Bitter, et al., Phys Rev A, 19 (1979) 1770-1779.
[7] E. Kallne, J. Kallne, J.E. Rice, Phys Rev Lett, 49 (1982) 330-333.
[8] E. Kallne, J. Kallne, R.D. Cowan, Phys Rev A, 27 (1983) 2682-2697.
[9] E. Kallne, J. Kallne, A. Dalgarno, et al., Phys Rev Lett, 52 (1984) 2245-2248.
[10] K.W. Hill, M. Bitter, M. Tavernier, et al., Rev Sci Instrum, 56 (1985) 848-848.
[11] E. Kallne, J. Kallne, E.S. Marmar, et al., Phys Scripta, 31 (1985) 551-564.
[12] J.E. Rice, E.S. Marmar, J.L. Terry, et al., Phys Rev Lett, 56 (1986) 50-53.
[13] J.E. Rice, E.S. Marmar, Rev Sci Instrum, 61 (1990) 2753-2755.
[14] J.E. Rice, F. Bombarda, M.A. Graf, et al., Rev Sci Instrum, 66 (1995) 752-754.
[15] J.E. Rice, K.B. Fournier, M.A. Graf, et al., Phys Rev A, 51 (1995) 3551-3559.
[16] J.E. Rice, K.B. Fournier, J.L. Terry, et al., Phys Rev A, 53 (1996) 3953-3962.
[17] J.E. Rice, J.L. Terry, K.B. Fournier, et al., J Phys B-at Mol Opt, 29 (1996) 2191-2208.
[18] J.E. Rice, J.L. Terry, J.A. Goetz, et al., Phys Plasmas, 4 (1997) 1605-1609.
[19] J.E. Rice, J.L. Terry, E.S. Marmar, et al., Nucl Fusion, 37 (1997) 241-249.
[20] M. Bitter, K.W. Hill, A.L. Roquemore, et al., Rev Sci Instrum, 70 (1999) 292-295.
[21] J.E. Rice, K.B. Fournier, J.A. Goetz, J Phys B-at Mol Opt, 33 (2000) 5435-5462.
[22] J.E. Rice, J.A. Goetz, R.S. Granetz, et al., Phys Plasmas, 7 (2000) 1825-1830.
[23] M. Bitter, K. Hill, L. Roquemore, et al., Rev Sci Instrum, 74 (2003) 1977-1981.
[24] A. Ince-Cushman, J.E. Rice, S.G. Lee, et al., Rev Sci Instrum, 77 (2006) 10F321.
[25] A. Ince-Cushman, J.E. Rice, M. Bitter, et al., Rev Sci Instrum, 79 (2008) 10E302.
[26] S.G. Lee, J.G. Bak, U.W. Nam, et al., Rev Sci Instrum, 81 (2010) 10E506.
[27] Yuejiang Shi, Fudi Wang, Baonian Wan, et al., Plasma Phys Contr F, 52 (2010) 085014.
[28] Fudi Wang, Yuejiang Shi, Wei Zhang, et al., Journal of Korean Physical Society, 59 (2011) 2734-2738.
[29] N.A. Pablant, M. Bitter, L. Delgado-Aparicio, et al., Rev Sci Instrum, 83 (2012) 083506.
[30] A. Langenberg, N.A. Pablant, T. Wegner, et al., Rev Sci Instrum, 89 (2018).
[31] D.M. Yao, G.N. Luo, Z.B. Zhou, et al., Phys Scripta, T167 (2016).
[32] G.S. Xu, L. Wang, D.M. Yao, et al., Nucl Fusion, 61 (2021).
[33] B.N. Wan, Y. Liang, X.Z. Gong, et al., Nucl Fusion, 59 (2019).
[34] B.N. Wan, Y.F. Liang, X.Z. Gong, et al., Nucl Fusion, 57 (2017).
[35] M.F. Gu, Astrophys J, 582 (2003) 1241-1250.
[36] M.F. Gu, Astrophys J, 593 (2003) 1249-1254.
[37] M.F. Gu, Astrophys J, 590 (2003) 1131-1140.
[38] M.F. Gu, Astrophys J, 589 (2003) 1085-1088.
[39] M.F. Gu, Can J Phys, 86 (2008) 675-689.
[40] H.-K. Chung, M. H. Chen, W.L.Morgan, et al., High Energ Dens Phys, 1 (2005) 3-12.
[41] R.W. Lee, J.T. Larsen, J Quant Spectrosc Ra, 56 (1996) 535-556.
[42] H.K. Chung, W.L. Morgan, R.W. Lee, J Quant Spectrosc Ra, 81 (2003) 107-115.
[43] J.E. Rice, M. Gu, N.M. Cao, et al., J Phys B-at Mol Opt, 54 (2021).
[44] B.N. Wan, J.G. Li, H.Y. Guo, et al., Nucl Fusion, 53 (2013).
[45] H.D. Xu, X.J. Wang, F.K. Liu, et al., Plasma Science & Technology, 18 (2016) 442-448.
[46] H.D. Xu, W.Y. Xu, D.J. Wu, et al., Fusion Eng Des, 164 (2021).

[47] W.Y. Xu, H.D. Xu, F.K. Liu, et al., Fusion Eng Des, 113 (2016) 119-125.
[48] L.M. Zhao, J.F. Shan, F.K. Liu, et al., Plasma Science & Technology, 12 (2010) 118-122.
[49] F.K. Liu, J.G. Li, J.F. Shan, et al., Fusion Eng Des, 113 (2016) 131-138.
[50] M. Wang, Z.G. Wu, W.D. Ma, et al., Ieee Access, 6 (2018) 37413-37417.
[51] W.D. Ma, M. Wang, Z.G. Wu, et al., Rev Sci Instrum, 90 (2019).
[52] Y.P. Zhao, X.J. Zhang, Y.Z. Mao, et al., Fusion Eng Des, 89 (2014) 2642-2646.
[53] C.D. Hu, Y.H. Xie, Y.L. Xie, et al., Plasma Science & Technology, 17 (2015) 817-825.
[54] C.D. Hu, Y.H. Xie, Y.L. Xie, et al., Rev Sci Instrum, 87 (2016).
[55] Y. Liu, H.L. Zhao, T.F. Zhou, et al., Fusion Eng Des, 136 (2018) 72-75.
[56] H.L. Zhao, T.F. Zhou, Y. Liu, et al., Rev Sci Instrum, 89 (2018).
[57] Q. Zang, J.Y. Zhao, L. Yang, et al., Rev Sci Instrum, 82 (2011).
[58] B. Lyu, F. D. Wang, X. Y. Pan, et al., Rev.Sci.Instru, 85 (2014) 11E406.
[59] F.D. Wang, J. Chen, R.J. Hu, et al., Rev Sci Instrum, 87 (2016).
[60] L. A. Vainshtein, U.I. Safronova, At. Data Nucl. Data Tables, 21 (1978) 49.
[61] T. Shirai, Y. Nakai, K. Ozawa, et al., J Phys Chem Ref Data, 16 (1987) 327-377.
[62] T. Shirai, J. Sugar, A. Musgrove, et al., J Phys Chem Ref Data, DOI (2000) 3-+.
[63] A.H. Gabriel, Mon. Not. R. Astron. Soc., 160 (1972) 99.
[64] N. Tragin, J.P. Geindre, P. Monier, et al., Phys Scripta, 37 (1988) 72-82.
[65] J. Clementson, P. Beiersdorfer, G.V. Brown, et al., Phys Scripta, 81 (2010).
[66] X.Y. Pan, F.D. Wang, J. Chen, et al., Fusion Eng Des, 96-97 (2015) 844-847.